\documentstyle[12pt]{article}
\def\be{\begin{equation}}
\def\ee{\end{equation}}
\def\bc{\begin{center}}
\def\ec{\end{center}}
\def\bea{\begin{eqnarray}}
\def\eea{\end{eqnarray}}
\def\dd{\displaystyle}
\def\nn{\nonumber}
\def\DD{{\cal D}}
\def\LL{{\cal L}}
\def\dmu{\partial_\mu}
\def\dnu{\partial_\nu}

\catcode`@=11
\def\marginnote#1{}
\newcount\hour
\newcount\minute
\newtoks\amorpm
\hour=\time\divide\hour by60
\minute=\time{\multiply\hour by60 \global\advance\minute by-\hour}
\edef\standardtime{{\ifnum\hour<12 \global\amorpm={am}%
        \else\global\amorpm={pm}\advance\hour by-12 \fi
        \ifnum\hour=0 \hour=12 \fi
        \number\hour:\ifnum\minute<10 0\fi\number\minute\the\amorpm}}
\edef\militarytime{\number\hour:\ifnum\minute<10 0\fi\number\minute}
\def\draftlabel#1{{\@bsphack\if@filesw {\let\thepage\relax
   \xdef\@gtempa{\write\@auxout{\string
      \newlabel{#1}{{\@currentlabel}{\thepage}}}}}\@gtempa
   \if@nobreak \ifvmode\nobreak\fi\fi\fi\@esphack}
        \gdef\@eqnlabel{#1}}
\def\@eqnlabel{}
\def\@vacuum{}
\def\draftmarginnote#1{\marginpar{\raggedright\scriptsize\tt#1}}
\def\draft{\oddsidemargin 0.0truein
        \def\@oddfoot{\sl preliminary draft \hfil
        \rm\thepage\hfil\sl\today\quad\militarytime}
        \let\@evenfoot\@oddfoot \overfullrule 3pt
        \let\label=\draftlabel
        \let\marginnote=\draftmarginnote
   \def\@eqnnum{(\theequation)\rlap{\kern\marginparsep\tt\@eqnlabel}%
\global\let\@eqnlabel\@vacuum}  }
\catcode`@=12
\begin{document}

\begin{titlepage}
\vspace*{-1cm}
\phantom{bla}
\hfill{DFPD 95/TH 27}
\vskip 1.5cm
\begin{center}
{\Large\bf Bounds on Heavy Chiral Fermions
}
\footnote{Work supported in part by the European Union
under contract No.~CHRX-CT92-0004.}
\end{center}
\vskip 1.0cm
\begin{center}
{\large Antonio Masiero}
\\
\vskip .1cm
Dipartimento di Fisica, Universit\'a di Perugia, Italy
\\
INFN, Sezione di Perugia, Italy
\\
\vskip .3cm
{\large Ferruccio Feruglio,  Stefano Rigolin}
\\
\vskip .1cm
Dipartimento di Fisica, Universit\'a di Padova, Italy
\\
INFN, Sezione di Padova, Padua, Italy
\\
\vskip .2cm
and
\\
\vskip .2cm
{\large Roberto Strocchi}
\\
\vskip .1cm
Dipartimento di Fisica, Universit\'a di Roma, Italy
\\
INFN, Sezione di Roma, Italy
\end{center}
\vskip 0.5cm
\begin{abstract}
\noindent
We derive
the low-energy electroweak effective lagrangian for the case
of additional heavy, unmixed, sequential fermions.
Present data still allow for the presence of a new quark and/or lepton
doublet with masses greater than $M_Z/2$, provided that these multiplets
are sufficiently degenerate.
Deviations of the effective lagrangian predictions from a full
one-loop computation are sizeable only for fermion masses
close to the threshold $M_Z/2$. Some of the constraints on new sequential
fermions coming from accelerator results and cosmological considerations are
presented. We point out that the new fermions can significantly affect
the production and decay rate into $\gamma \gamma$ of the intermediate
Higgs at LHC.
\end{abstract}
\vfill{
May 1995}
\end{titlepage}
\setcounter{footnote}{0}
\vspace{1cm}
LEP precision data represent a step of paramount
relevance in probing extensions of the Standard Model (SM).
Through their virtual effects, the electroweak radiative corrections "feel"
the presence of new particles running in the loops and the level of
accuracy on the relevant observables is such that this set of tests is
complementary to the traditional probes on virtual effects due to new
physics (i.e. highly suppressed or forbidden flavour changing neutral
current phenomena). In some cases, as that which we aim to discuss here,
the electroweak precision tests represent the only indirect way to
search for these new particles.

In this letter we will discuss electroweak radiative effects
from extensions of the ordinary fermionic spectrum of the SM.
The new fermions are supposed to
possess the same colour and electroweak quantum numbers as the
ordinary ones and to
mix very tinily with the ordinary three generations.

The most straightforward realization of such a fermionic extension
of the SM is the introduction of a fourth generation of
fermions. This possibility has been almost entirely jeopardized
by the LEP bound on the numbers of neutrinos species. Although
there still exists the obvious way out of having new fermion
generations with neutrinos of mass $ \ge M_Z/2 $, we think
that these options are awkward enough not to deserve further studies.
Rather, what we have in mind in tackling this problem are
general frames discussing new physics beyond the SM
which lead to new quarks and/or leptons classified in
the usual chiral way with iso-doublets and iso-singolets for
different chiralities.

Situations of this kind may be encountered in grand
unified schemes where the ordinary fifteen Weyl spinors of each
fermionic generation are only part of larger representations or
where new fermions (possibly also mirror fermions) are requested
by the group or manifold structure of the schemes.
Chiral fermions with heavy static masses may also provide
a first approximation of virtual effects in techicolor-like
schemes when the dynamical behaviour of
technifermion self-energies are neglected.

Although such effects have been extensively investigated in the literature
\cite{gen}, our presentation focuses mainly on two aspects, which
have been only partially touched in the previous analyses: the use
of effective lagrangians for a model-independent treatment of the problem
and a discussion of the validity of this approach in comparison with the
computation in the full-fledged theory.

While separate tests can be set up for each different extension of the SM,
there may be
some advantage in realizing this analysis in a model independent
framework. The natural theoretical tool to this purpose
is represented by an effective electroweak lagrangian where,
giving up the renormalizability
requirement, all $SU(2)_L\otimes U(1)_Y$ invariant operators up to a given
dimension are present with unknown coefficients, to
be eventually determined
from the experiments. Each different model fixes uniquely this set of
coefficients and the effective lagrangian becomes in this way a common
ground to discuss and compare several SM extensions.
The introduction of the well known $S$, $T$ and $U$ \cite{stu} or
$\epsilon$'s \cite{eps}
variables was much in the same spirit and the use of an effective lagrangian
represents in a sense the natural extension of these approaches.

The use of an effective lagrangian for the electroweak physics
has been originally advocated for the study of the large Higgs mass limit
in the SM \cite{abe,alo,her}. Subsequently,
contributions from chiral $SU(2)_L$ doublets have
been considered in the degenerate case
\cite{dho},
for small splitting \cite{app} and in the case of infinite splitting
\cite{ste,fmm}.
In the present note we will deal with the general case
of arbitrary splitting among the fermions in the doublet.
Our results will be used to test
the model with the latest available data.

The use of an effective lagrangian in precision tests has its own
limitations, which are also discussed in the present note.
In particular we are going to use an effective electroweak lagrangian
organized in a derivative expansion which we truncate at the fourth
order. When discussing two-point vector boson functions
$-i\Pi^{\mu\nu}_{ij}(q)~~~(i,j=0,1,2,3)$, this amounts
to keep only the constant and linear terms in $q^2$:
\be
\Pi^{\mu\nu}_{ij} (q) = \Pi_{ij}(q^2) g^{\mu\nu} + (q^\mu q^\nu ~~{\rm terms})
\label{a0}
\ee
\be
\Pi_{ij}(q^2)\equiv A_{ij} + q^2 F_{ij}(q^2) =
             A_{ij} + q^2 F_{ij}(0) +...
\label{a1}
\ee

The next terms in the $q^2$ expansion are suppressed by increasing
powers of $q^2/M^2$, $M$ generically representing the mass of the particles
running in the loop. One can ask how large has to be $M$
to obtain a sensible approximation from the truncation of the full one-loop
result. As we will see, for LEP I observables, the truncation is
a very good approximation already for relatively light fermions,
with masses around $70-80~GeV$.

At the end of this note we will add some comments on the direct searches
of new quarks and leptons and on the modifications induced by additional
chiral fermions in the $\gamma\gamma$ signature for an intermediate
Higgs at LHC.

\vspace{1cm}
For new chiral fermions which do not mix with the ordinary ones,
the virtual effects measurable at LEP 1 are all described by operators
bilinear in the gauge vector bosons.
Here, for completeness, we consider the standard list \cite{alo}
of CP conserving
operators containing up to four derivatives and built out of
the gauge vector bosons $W^i_\mu~~(i=1,2,3),~B_\mu$
and the would be Goldstone bosons $\xi^i$:
\bea
\LL_0&=&\dd\frac{v^2}{4}[tr(TV_\mu)]^2\nn\\
\LL_1&=&i\dd\frac{gg'}{2}B_{\mu\nu}tr(T\hat W^{\mu\nu})\nn\\
\LL_2&=&i\dd\frac{g'}{2}B_{\mu\nu}tr(T[V^\mu,V^\nu])\nn\\
\LL_3&=&gtr(\hat W_{\mu\nu}[V^\mu,V^\nu])\nn\\
\LL_4&=&[tr(V_\mu V_\nu)]^2\nn\\
\LL_5&=&[tr(V_\mu V^\mu)]^2\nn\\
\LL_6&=&tr(V_\mu V_\nu)tr(TV^\mu)tr(TV^\nu)\nn\\
\LL_7&=&tr(V_\mu V^\mu)[tr(TV^\nu)]^2\nn\\
\LL_8&=&\dd\frac{g^2}{4}[tr(T\hat W_{\mu\nu})]^2\nn\\
\LL_9&=&\dd\frac{g}{2}tr(T\hat W_{\mu\nu})tr(T[V^\mu,V^\nu])\nn\\
\LL_{10}&=&[tr(TV_\mu)tr(TV_\nu)]^2\nn\\
\LL_{11}&=&tr((\DD_\mu V^\mu)^2)\nn\\
\LL_{12}&=&tr(T\DD_\mu \DD_\nu V^\nu)tr(TV^\mu)\nn\\
\LL_{13}&=&\dd\frac{1}{2}[tr(T \DD_\mu V_\nu)]^2\nn\\
\LL_{14}&=&ig\epsilon^{\mu\nu\rho\sigma}tr(\hat W_{\mu\nu}V_\rho)tr(TV_\sigma)
\label{a2}
\eea
We recall the notation:
\bea
T&=&U\tau^3U^\dagger~~~,\nn\\
V_\mu&=&(D_\mu U) U^\dagger~~~,
\label{a3}
\eea
\be
U=e^{\dd i\frac{\vec\xi\cdot\vec\tau}{v}}~~~,
\label{a4}
\ee
\be
D_\mu U=\dmu U - g \hat W_\mu U+g'U\hat B_\mu~~~,
\label{a5}
\ee
$\hat W_\mu,~\hat B_\mu$ are matrices collecting the gauge fields:
\bea
\hat W_\mu&=&\frac{1}{2i}\vec W_\mu\cdot\vec\tau~~~,\nn\\
\hat B_\mu&=&\frac{1}{2i} B_\mu\tau^3~~~.
\label{a6}
\eea
The corresponding field strengths are given by:
\bea
\hat W_{\mu\nu}&=&\dmu\hat W_\nu-\dnu\hat W_\mu-g[\hat W_\mu,\hat W_\nu]
{}~~~,\nn\\
\hat B_{\mu\nu}&=&\dmu\hat B_\nu-\dnu\hat B_\mu~~~.
\label{a7}
\eea
Finally the covariant derivative acting on $V_\mu$ is given by:
\be
\DD_\mu V_\nu=\dmu V_\nu-g[\hat W_\mu,V_\nu]~~~.
\label{a8}
\ee
The effective electroweak lagrangian reads:
\be
\LL_{eff}=\LL_{SM} + \sum_{i=0}^{14} a_i \LL_i~~~,
\label{a9}
\ee
where $\LL_{SM}$ is the SM lagrangian. Here we do not include
the Wess-Zumino term \cite{fmm}.
For an extra doublet of fermions (quarks or leptons), we have determined
the coefficients $a_i~(i=0,...14)$, by computing
the corresponding one-loop contribution to
a set of $n$-point gauge boson functions $(n=2,3,4)$,
in the limit of low external momenta and by matching the predictions of
the full and effective theories.
By denoting with $M$ and $m$ the masses of the upper and lower
weak isospin components, respectively, we obtain, in units of $1/16\pi^2$:
\bea
a^q_0  &=&  {3 M^2 \over 2 v^2}\left( { 1 - r^2 + 2\,r\,\log r
\over 1-r} \right)
\nn\\
a^q_1  &=& \frac{1}{12 (-1+r)^3}\left[ {3(1 - 15 r + 15 r^2 - r^3) + 2 (1 -
12 r  -6 r^2  - r^3 )\log r } \right]
\nn\\
a^q_2 &=& \frac{1}{12 (-1+r)^3}\left[{3(3 - 7\,r + 5\,{r^2} - r^3)+ 2\,(1 -
{r^3})\,\log r } \right]
\nn\\
a^q_3 &=& \frac{1}{8 (-1+r)^3}\left[ 3 ( -1 + 7\,r - 7\,{r^2} +
{r^3}) + 6\,r\,(1 + {r})\,\log r) \right]
\nn\\
a^q_4 &=& \frac{1}{6 (-1+r)^3}\left[ 5 - 9\,r + 9\,{r^2} - 5\,{r^3} +
3\,(1 + {r^3})\,\log r \right]
\nn\\
a^q_5 &=& \frac{1}{24 (-1+r)^3}\left[ -23 + 45\,r - 45\,{r^2} + 23\,{r^3}
- 12\,(1+{r^3})\,\log r \right]
\nn\\
a^q_6  &=& \frac{1}{24 (-1+r)^3}
\left[ -23 + 81\,r - 81\,r^2 + 23\,r^3
-6\,(2 -3\,r\, -3\,r^2\,
+2\,r^3)\,\log r \right]
\nn\\
a^q_7  &=&-a^q_6
\nn\\
a^q_8 &=&\frac{1}{12 (-1+r)^3}\left[ 7 - 81\,r + 81\,r^2 - 7\,r^3
+ 6\,(1 - 6\,r\, - 6\,r^2\, + r^3)\,\log r \right]
\nn\\
a^q_9 &=&-a^q_6
\nn\\
a^q_{10}&=&0
\nn\\
a^q_{11}&=&-{1\over{2}}
\nn\\
a^q_{12}&=&\frac{1}{8 (-1+r)^3}
\left[1 + 9\,r - 9\,{r^2} - {r^3} + 6\,r\, (1+
{r})\,\log r\right]
\nn\\
a^q_{13} &=& 2\,a^q_{12}
\nn\\
a^q_{14} &=&\frac{3}{8 (-1+r)^2}\left[ 1 - r^2 + 2 r \log r
\right]
\label{b1}
\eea
for quarks and:
\bea
a^l_i  &=&\frac{1}{3} a^q_i~~~~~~~~~(i=0,~~i=3,...14)
\nn\\
a^l_1  &=&  \frac{1}{ 12 (-1+ r)^3}
\left[ 1 - 15  r + 15  r^2 -  r^3 - 2 (1 +
6  r^2  -   r^3 )\log  r  \right]
\nn\\
a^l_2 &=&\frac{1}{ 12 (-1+ r)^3}
\left[-1 - 3\, r + 9\,{ r^2} - 5\, r^3- 2\,(1 -
{ r^3})\,\log  r  \right]
\label{b2}
\eea
for leptons, where:
\be
r=\frac{m^2}{M^2}
\label{a10}
\ee
The coefficients $a_i$ of the effective lagrangian $\LL_{eff}$ are related
to measurable parameters. In particular,
to make contact with the LEP data, we recall that, by neglecting
higher derivatives, the relation between the effective lagrangian
$\LL_{eff}$ and the $\epsilon$'s parameters, is given by:
\bea
\delta\epsilon_1&=&2 a_0~,\nn\\
\delta\epsilon_2&=&-g^2(a_8+a_{13})~,\nn\\
\delta\epsilon_3&=&-g^2(a_1+a_{13})~.
\label{a11}
\eea
The $\epsilon$ parameters are obtained by adding to $\delta\epsilon_i$
the SM contribution $\epsilon^{SM}_i$, which we regard as functions of
the Higgs and top quark masses.
{}From eqs. (\ref{b1}) and (\ref{b2}) one finds:
\be
\delta\epsilon^q_1=3 \delta\epsilon^l_1=
{3 M^2 \over 8 \pi^2 }{G \over \sqrt{2}} \left[ { 1 - r^2
+ 2\,r\,\log r \over (1-r)} \right]
\label{a12}
\ee
\be
\delta\epsilon^l_2=3 \delta\epsilon^l_2=
 {G m_W^2 \over 12 \pi^2 \sqrt{2}} \left[ {5 - 27r + 27
r^2-5 r^3+\left(3 - 9 r - 9 r^2 +3 r^3 \right) \log r \over \left( 1 - r
\right)^3} \right]
\label{a13}
\ee
\be
\delta\epsilon^q_3 = {G m_W^2 \over 12 \pi^2 \sqrt{2}} \left[ 3 +  \log r
\right]
\label{a14}
\ee
\be
\delta\epsilon^l_3 = {G m_W^2 \over 12 \pi^2 \sqrt{2}} \left[ 1 - \log r
\right]
\label{a15}
\ee
A recent analysis of the available precision data from LEP, SLD,
low-energy neutrino scatterings and atomic parity violation experiments,
leads to the following values for the $\epsilon$ parameters
\cite{fit}:
\bea
\epsilon_1&=&(3.6\pm 1.5)\cdot 10^{-3}\nn\\
\epsilon_2&=&(-5.8\pm 4.3)\cdot 10^{-3}\nn\\
\epsilon_3&=&(3.6\pm 1.5)\cdot 10^{-3}
\label{eps}
\eea
Notice the relatively large error in the determination of $\epsilon_2$,
mainly dominated by the uncertainty on the $W$ mass.
We illustrate our result in fig. 1, in the plane $(\epsilon_1,\epsilon_3)$,
for the case of an extra quark doublet. The upper ellipsis represents
the 1 $\sigma$ experimentally allowed region, obtained by combining all
LEP data.

If one also includes the SLD determination of the left-right
asymmetry, then one gets the lower ellipsis. The predictions from an
additional heavy quark doublet are given by the dashed line,
obtained by fixing one of the masses to $200~GeV$ and letting the other
vary from $200~GeV$ to $300~GeV$. One has in this way two branches,
according to which mass, $m$ or $M$, has been fixed. The top and Higgs
masses has been fixed to $175~GeV$ and $100~GeV$, respectively.
As expected, it appears that only a small amount of splitting
among the doublet components is allowed. For the chosen value of $m_{t}$
and $m_H$, the SM prediction lies already outside the
1 $\sigma$ allowed region and additional positive contributions to
$\epsilon_1$ tend to be disfavoured. On the contrary, the positive
contribution to $\epsilon_3$, almost constant in the chosen range
of masses, is still tolerated, and even preferred by the fit to
the data which do not include the SLD result.

The result for a full extra generation of heavy quarks and leptons
is shown in fig. 2, dashed line. We have assumed equal ratio $r$
in the lepton and in the quark sector. As can be seen from
eqs. (\ref{a14}-\ref{a15}), this makes the two branches of fig. 1
to degenerate in a unique line.

If new physics beyond the SM were modeled by
additional heavy chiral fermions,
of the kind we have considered, then, from the effective lagrangian
of eqs. (\ref{a9}-\ref{a10}) we could draw informations on
the future searches of anomalous trilinear couplings.
Indeed,
the anomalous magnetic and weak moments of the $W$, $\Delta k_\gamma$
and $\Delta k_Z$, can be expressed as combinations of the coefficients
$a_i$. One finds \cite{par,app}\footnote{
The definitions of the anomalous couplings depend on the overall
normalization of the trilinear $WWN~~~(N=\gamma,Z)$ vertex, usually
denoted by $g_{WWN}$. Here we are following the convention
of ref. \cite{app}.}:
\bea
\Delta k_\gamma&=&g^2(-a_1+a_2-a_3+a_8-a_9)\nn\\
\Delta k_Z&=&\frac{a_0}{(c^2-s^2)}+\frac{g^2 s^2}{(c^2-s^2) c^2} (a_1+a_{13})
\nn\\
          &+&g^2[\frac{s^2}{c^2}(a_1+a_{13}-a_2)-a_3+a_8-a_9+a_{13}]
\label{a15bis}
\eea
$s$ and $c$ denoting the sine and the cosine of the Weinberg angle.
The contribution of a quark or lepton doublet to the anomalous moments
can be readily evaluated by substituting in eq. (\ref{a15bis}) the explicit
expressions of the coefficients $a_i$ given in eqs. (\ref{a9}-\ref{a10}).
For instance, for the anomalous magnetic moment $\Delta k_\gamma$, we obtain:
\bea
\Delta k_\gamma^q&=&\frac{G m_W^2}{4 \pi^2 \sqrt{2}}\frac{1}{(1-r)^3}
[-1+8 r-7 r^2 + 2 (r+2 r) \log r]~~~~,\\
\Delta k_\gamma^l&=&\frac{G m_W^2}{12 \pi^2 \sqrt{2}}\frac{1}{(1-r)^3}
[1+6 r-9 r^2 + 2 r^3 + 6 r \log r]
\eea
for quarks and leptons, respectively. Considering in particular the case
of degenerate doublets, as suggested by the small value of the $\epsilon_1$
parameter, one finds:
\be
\Delta k_\gamma^q=3\Delta k_\gamma^l=-\frac{G m_W^2}{4 \pi^2 \sqrt{2}}\sim
- 1.3\cdot 10^{-3}~~~~~.
\ee
Similar contributions are also obtained for
$\Delta k_Z$ so that
only very large multiplicities (many doublets) could push the predictions for
the anomalous couplings to the level of observability available
at LEP II. As a consequence,
the contributions to $\epsilon_3$ would be necessarily positive,
large and hard to reconcile with the data.

We are thus lead to consider the possibility of relatively
light $(m,M\ge M_Z/2)$ chiral fermions, both to check the agreement
with the present data, and to test the reliability
of our effective lagrangian approach. If the additional fermions
are not sufficiently heavy, we do not expect that their one-loop
effects are accurately reproduced by the coefficients $a_i$ in
eqs. (\ref{b1}-\ref{b2}). In this case we have to consider the full dependence
on external momenta
of the Green functions, not just the first two terms of the
$q^2$ expansion given in eq. (\ref{a1}).
We recall that in this case the $\epsilon$
parameters are given by  \cite{bfc}:
\bea
\delta\epsilon_1&=&e_1-e_5\nn\\
\delta\epsilon_2&=&e_2-s^2 e_4 -c^2 e_5\nn\\
\delta\epsilon_1&=&e_3+c^2 e_4-c^2 e_5
\label{a16}
\eea
where we have kept into account the fact that in our case
there are no vertex or box corrections to four-fermion processes.
In eq. (\ref{a16})
\bea
e_1&=&\frac{A_{33}-A_{WW}}{M_W^2}\nn\\
e_2&=&F_{WW}(M_W^2)-F_{33}(M_Z^2)\nn\\
e_3&=&\frac{c}{s} F_{30}(M_Z^2)\nn\\
e_4&=&F_{\gamma\gamma}(0)-F_{\gamma\gamma}(M_Z^2)\nn\\
e_5&=&M_Z^2 F'_{ZZ}(M_Z^2)
\label{a17}
\eea
where the quantities $A_{ij}$ and $F_{ij}$ are defined in
eqs. (\ref{a0}) and (\ref{a1}).
The expressions for the quantities $e_i$, in the case of
an ordinary quark or lepton doublet can be easily derived from
the literature \cite{yel}. We plot the result for an extra quark doublet
in fig. 1 (full line). We have taken $m_{t}=175~GeV$ and
$m_H=100~GeV$. One of the two masses is kept fixed at $50~GeV$,
and the other one runs from $50~GeV$ to $170~GeV$.
The small masses cause a substantial deviation from the
asymptotic, effective lagrangian prediction. Nevertheless,
as illustrated in fig. 1, such a large effect is still compatible with
the present data. The case of a full new generation is shown in the
solid line of fig. 2.


In particular, as it was observed in \cite{bfc},
a large negative contribution to both $\epsilon_1$ and $\epsilon_3$
is now possible, due to a formal divergence of $F'_{ZZ}$
at the threshold which produces a large and positive $e_5$.
Clearly, this behaviour cannot be reproduced by $\LL_{eff}$,
which, at the fourth order in derivatives, automatically sets
$F'_{ZZ}=0$. A relevant question is, then, when the asymptotical regime
starts, i.e. how close to $M_Z$ should be the masses of the new quarks or
leptons for observing deviations due to the full expression of $\Pi_{ij}(q^2)$
instead of the truncated expression given in
eq. (\ref{a1}). A detailed analysis shows
that already for masses of the new fermions above $70 - 80 \ GeV$ the
difference between the values of the $\epsilon_i$ obtained with the truncated
and full expression of $\Pi_{ij}(q^2)$ are as small as $10^{-4}$,
i.e. below
the present experimental level of accuracy. This is illustrated in fig. 3
where the asymptotical and full expression of $\epsilon_3$ are
compared as a function of $r$.
\vspace{1cm}

Beyond the indirect precision tests,
the possibility of having new fermions carrying the usual $SU(3)_C
\times SU(2)_L \times U(1)_Y$ quantum numbers can clearly also be
bounded by the direct searches.

Concerning the present searches,
from LEP we have the lower bound of $M_Z/2$ which applies
independently from any assumption on the decay modes of the new fermions
which couple to the $Z$ boson. Much stronger limits on the new quarks masses
can be inferred from the Tevatron results. However, as we know from the
search for the top quark, these latter bounds rely on assumptions concerning
the decay modes of the heavy quark. For instance, in the case of the top
search it was stressed that if a new decay channel into the $b$ quark and a
charged Higgs were avaible to the top, then one could not use the CDF
bounds on $m_t$ \cite{cdf} which came along these last years, before the
final discovery of the top quark.

Now, it may be conceivable that the new physics related to the presence of
extra-fermions can also affect their possible decay channels making the
lightest of the new fermions unstable.
Indeed, we stated in our assumption that the new fermions do not
essentially mix with the ordinary ones, hence one has to invoke new physics
if one wants to avoid the formation of stable heavy mesons made out of the
lightest stable new fermion and of the ordinary fermions of the Standard
Model. If the new fermions can decay within the detector,
then the bounds on their
masses, coming from Tevatron data, must be discussed in a model-dependent
way and even the case of new quarks with masses close to $M_Z/2$
is not fully ruled out.


If on the contrary the lightest new quark is stable, then searches for exotic
heavy meson at CDF already ruled out the possibility of being near the
threshold $M_Z/2$. The existence of coloured particle with charge $\pm 1$
is strictly bounded over $130 \ GeV$ from CDF experiment \cite{sta}.
Finally, note that for charged leptons the bound coming from CDF are much
less stringent. A new stable charged lepton of mass of $50-60 \ GeV$ cannot
be ruled out.

We are aware of the fact that apart from bounds coming from the direct
production of the new fermions, there exist also cosmological limits which
apply to the case of stable electrically charged and (or) coloured fermions.
Cosmologically stable quarks of masses up to $20 \ TeV$ can be ruled out on
the basis of the results concerning superheavy element searches \cite{cos}.
As for stable charged hadronic superheavies in the $20-10^{5} \ TeV$ range
they seem
to be in contrast with the bounds which are obtained requiring that heavy
particles captured by neutrons star do not induce their collapse to a black
hole. Clearly all these severe bounds apply only to the case of
cosmologically stable new fermions, so that one can easily avoid them for
fermions which have some (even very small) mixing with the ordinary fermions
and (or) decay through particles which are related to the new physics beyond
the SM.

Finally, concerning future searches, we comment on an effect due to the
presence of new heavy quarks which may be potentially relevant for
the LHC physics.
We consider the production and decay into a photon pair of a Higgs of
intermediate mass, i.e. $100 \le m_H \le 150 \ GeV$. The presence of new
quarks give rise to competing effects of opposite sign at the level of
production and $\gamma \gamma$ decay of this intermediate Higgs.

Indeed, as
for production, the gluon-fusion amplitude increases due to the effect of the
new quarks which adds up to that of the top \cite{ggh}.
For a new doublet of quarks
heavier than the Higgs, the SM gluon-fusion amplitude gets approximately
multiplied by a factor 3, the total number of heavy quarks in the loop.
This leads to an
enhancement of a factor nine, roughly, for the cross section production.

However, in the decay into two photons the new quarks tend to decrease the
rate. The dominant contribution to $H \to \gamma \gamma$ comes from the loop
where the $W$ boson run, while the top exchange contribution yields an
opposite sign. Again the new heavy fermions produce contribution which are
analogous to those due to the top exchange and, hence, tend to reduce the
decay rate. More precisely, when additional fermions are present,
the partial width of the Higgs into a photon pair is given by \cite{hgg}:
\be
\Gamma(H\to\gamma\gamma)=\frac{\alpha^2}{128 \pi^3 \sqrt{2}}
G_F m_H^3 \left| F_1(\tau) +  \sum_{i} N_i Q_i^2 F_{1/2}
(\tau_i)\right|^2
\label{wid}
\ee
where the sum extends over all fermions of masses $M_i$,
$\tau=4 M_W^2/m_H^2$, $\tau_i= 4 M_i^2/m_H^2$,
$N_i=3$ for quarks and $N_i=1$ for leptons.
In the asymptotic regime, for $2 M_W > m_H$ and $2 M_i > m_H$, one has:
\be
F_1(\tau)= 7,~~~~~~~F_{1/2}(\tau_i)=-\frac{4}{3}~~~,
\label{tau}
\ee
When $m_H\simeq 100~GeV$, from eqs. (\ref{wid}) and (\ref{tau}),
one obtains, roughly:
\be
 \frac{\left[BR(H\to\gamma\gamma)\right]_{new}}
{\left[BR(H\to\gamma\gamma)\right]_{SM}}
\simeq {1 \over 3}
 \label{br1}
\ee
when only a new doublet of heavy quarks is present, and
\be
 \frac{\left[BR(H\to\gamma\gamma)\right]_{new}}
{\left[BR(H\to\gamma\gamma)\right]_{SM}}
\simeq {1 \over 10}
 \label{br2}
\ee
when a new complete generation of heavy quarks and leptons is considered.
Notice that the above
estimates for the $BR$'s are reliable as long as $m_H \simeq 100~GeV$.
For $m_H$ approaching $150~GeV$ the ratio $\tau$ gets
close to one and, thus the asymptotic expression of $F_1$ for the $W$ exchange
contribution given in eq. (\ref{tau}) can no longer be used.
In this latter case, the $W$ contribution
can become sizeably larger than in the asymptotic situation, hence making the
suppression of $BR(H \to \gamma \gamma)$ due to the new fermions less severe.
In the less favorable case of a full new heavy generation,
the product of the production cross-section times
the $BR(H \to \gamma \gamma)$ reflects a conspicuous dependence on $m_H$.
For $m_H\simeq 100~ GeV$ we obtain:
\be
\frac{\left[\sigma(gg\to H) \cdot BR(H\to\gamma\gamma)\right]_{new}}
{\left[\sigma(gg\to H) \cdot BR(H\to\gamma\gamma)\right]_{SM}}
\simeq 0.9
\label{ra1}
\ee
while for $m_H=150~GeV$ the above ratio becomes larger than one, reaching
values close to 4.

\vspace{1cm}

In conclusion, we discussed the impact of the presence of new sequential
fermions on the electroweak precision tests. We showed that the present data
still allowed the presence of a new quark and/or lepton doublet with masses
greater than $M_Z/2$.
Only for light new fermions which are close to the threshold $M_Z/2$ one finds
drastic departures of the effective lagrangian result
from the full one-loop radiative corrections obtained in
the SM. The presence of new fermions carrying usual $SU(3)_C
\times SU(2)_L \times U(1)_Y$ quantum numbers with mass as low as $60-80 \
GeV$ is severely limitated both by accelerator results and cosmological
constraints. Finally, the new fermions can significantly affect the production
and decay rate into $\gamma \gamma$ of the intermediate Higgs at LHC.
\vfill{
\section*{Acknowledgements}
We would like to thank L. Maiani for his partecipation in the early
stage of this work and for useful comments and suggestions.
We are indebted to M. Mangano and E. Nardi for valuable discussions
and to G. Altarelli, R. Barbieri and F. Caravaglios for
having promptly provided us with the latest $\epsilon$'s results.
}
\newpage
\section*{Figure Captions}

Fig. 1
Predictions for $\epsilon_1$, $\epsilon_3$ from an additional
quark doublet. The lower (upper)
dashed line represents the case $m~(M)~=200~GeV$,
$M~(m)$ varying between $200~GeV$ and $300~GeV$, evaluated
with ${\LL}_{eff}$. The lower (upper) full line corresponds
to $m~(M)~=50~GeV$,
$M~(m)$ varying between $50~GeV$ and $170~GeV$, evaluated
with a complete 1-loop computation. The SM point corresponds
to $m_t=175~GeV$ and $m_H=100~GeV$. The upper (lower) ellipses
is the 1 $\sigma$ allowed region, obtained by a fit of the high
energy data which excludes
(includes) the SLD measurement.

\noindent
Fig. 2
Same as for Fig. 1, in the case of a full extra generation of
quarks and leptons.

\noindent
Fig. 3
Comparison between the asymptotic (solid line) and full one-loop (dashed lines)
computations of $\epsilon_3$ versus $r=m^2/M^2$, for an additional
quark doublet. For the SM contribution, $m_t=175~GeV$ and $m_H=100~GeV$
are assumed. The 2 $\sigma$ allowed region is also displayed.
\vfill

\newpage

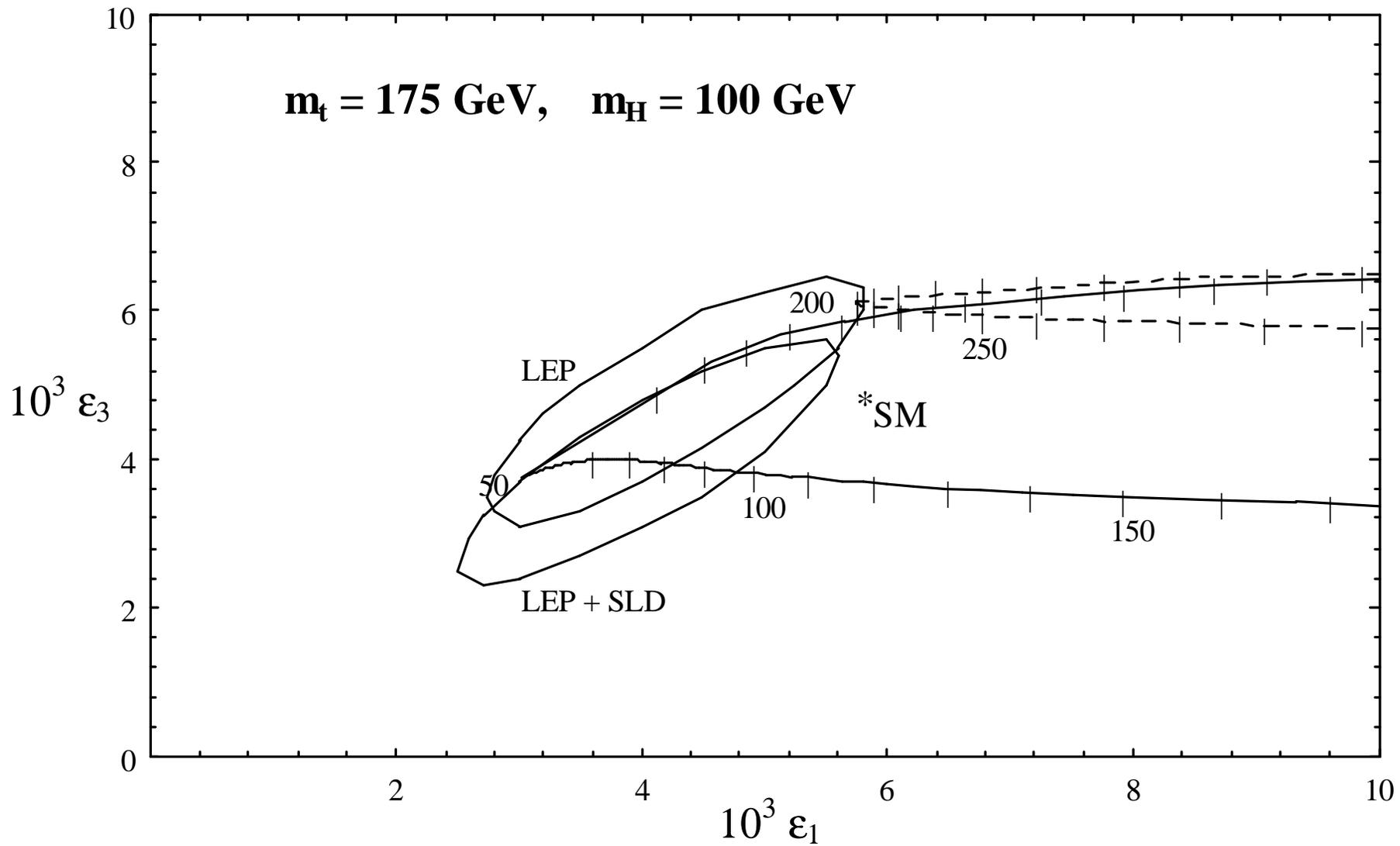
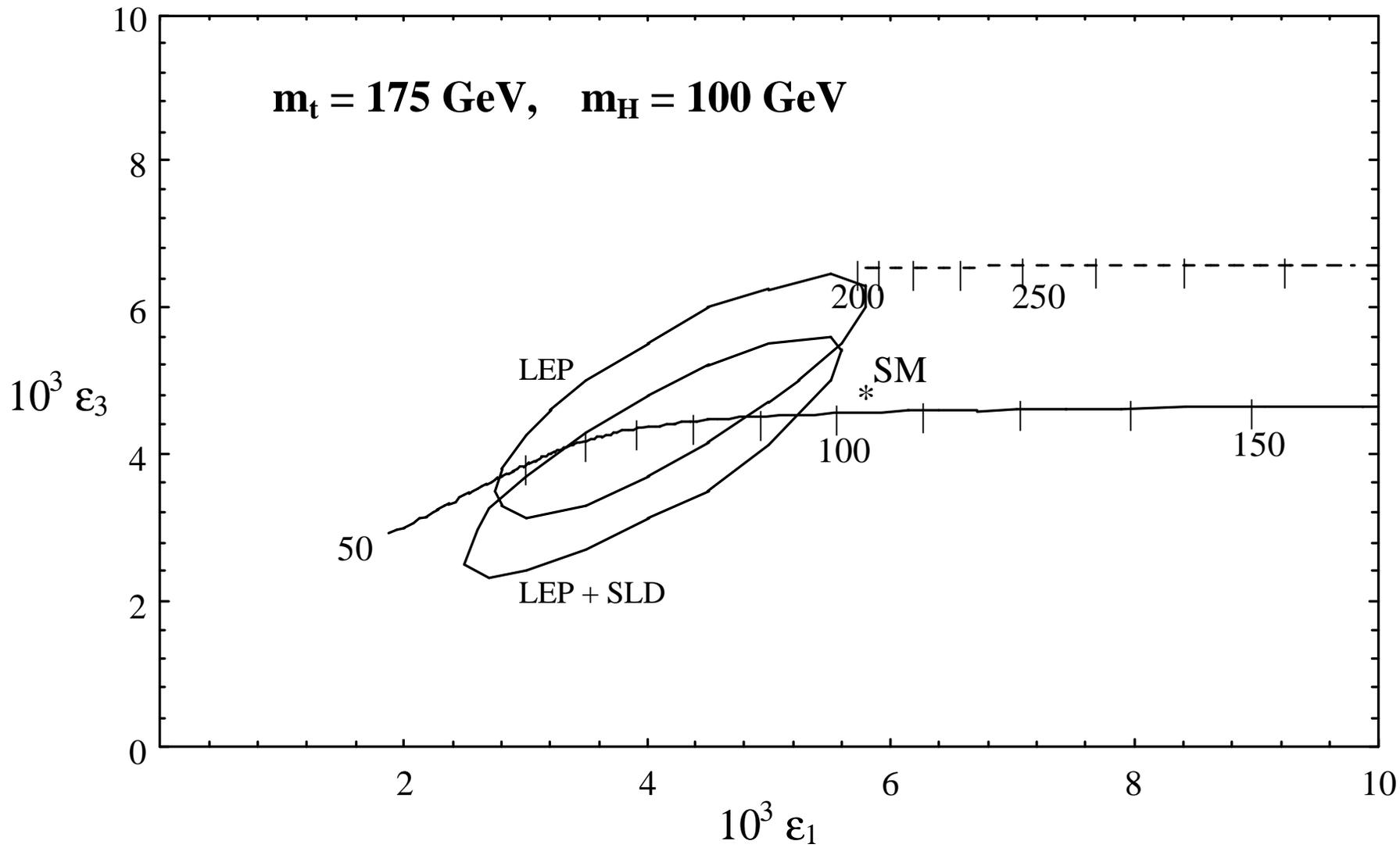
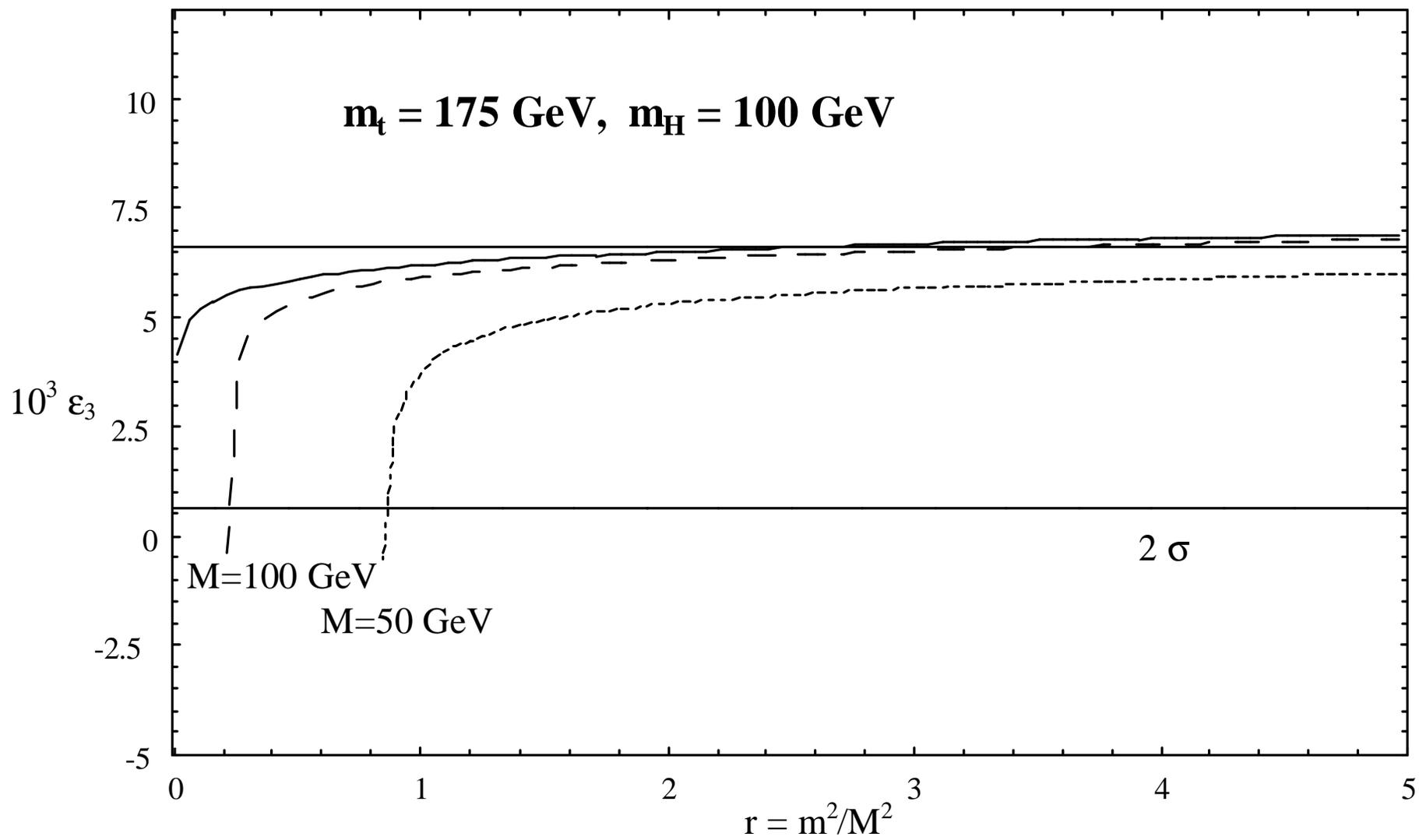
\end{document}